\documentclass[12pt]{article}
\usepackage[a4paper,margin=23mm]{geometry}
\usepackage[utf8]{inputenc}
\usepackage{amsmath,amssymb}
\usepackage{graphicx}
\usepackage{cite}
\usepackage{xurl}
\usepackage[section]{placeins}
\usepackage{caption}
\usepackage[hidelinks]{hyperref}


\newcommand{\rhot}{\rho_{\mathrm{tor}}}
\newcommand{\tN}{\widehat{t}}
\newcommand{\GT}{G_{T_i}}
\newcommand{\VZ}{\widehat{V}_{E,y}^{\mathrm{ZF}}}
\newcommand{\wZ}{\widehat{\omega}_{E\times B}}
\newcommand{\pZ}{\widehat{\varphi}_{Z}}
\newcommand{\avg}[1]{\langle #1\rangle_t}
\newcommand{\Ii}{\mathcal{I}_i}
\newcommand{\ExB}{E\times B}

\title{Fast-ion effects on \texorpdfstring{$E\times B$}{E x B} staircase organization and the heat-flux response to the ion-temperature gradient}

\author{Il-Hwan Kim and Jaemin Seo\textsuperscript{*}\\[0.35em]
\small Department of Physics, Chung-Ang University,\\
\small Seoul 06974, Republic of Korea\\[0.25em]
\small E-mail: \href{mailto:caih@cau.ac.kr}{caih@cau.ac.kr} (I.-H. Kim)\\
\small \textsuperscript{*}Corresponding author: \href{mailto:jseo@cau.ac.kr}{jseo@cau.ac.kr} (J. Seo)}

\date{}

\begin{document}

\maketitle

\begin{abstract}
Zonal-flow organization can modify the relation between heat flux and temperature gradient, but how fast ions affect this organization and the accompanying transport response remains less well characterized. We use nonlinear, global gyrokinetic simulations based on KSTAR fast-ion-regulated enhancement (FIRE) mode to compare cases with and without fast ions during profile relaxation across five initial main-ion temperature gradients. We observe that fast ions induce larger zonal-flow and shearing-rate amplitudes, together with persistent shear layers. At the reference profile based on the experiment, these shear layers coexist with temperature-gradient corrugations qualitatively consistent with $E\times B$ staircases. When fast ions are present, inward and outward ion heat fluxes coexist at different radii, which causes the average heat flux to be smaller than when fast ions are absent. This redistribution of ion heat transport may be relevant to understanding the improved ion confinement observed in FIRE mode. In particular, we note a transition-like response in the presence of fast ions, whereby the heat flux decreases as the ion-temperature gradient increases, in contrast to the conventional Fick's law trend without fast ions. This behavior that deviates from Fick's law suggests that FIRE mode may represent a bifurcation to a distinct transport regime rather than a continuous high-performance extension of L-mode. These findings highlight the importance of the initial temperature gradient and global zonal-flow organization when assessing fast-ion effects on ion heat transport.
\end{abstract}


\section{Introduction}

Microturbulent heat transport is a major constraint on energy confinement in magnetically confined plasmas in a tokamak. Drift-wave turbulence can drive zonal $\ExB$ flows through nonlinear interactions\cite{r1,jovanovic2001, smolyakov2000}. The resulting flow shear can suppress turbulence and reduce the associated heat transport\cite{r1,r2,r3}. Under certain conditions, multiple shear layers organize radially together with local corrugations in the temperature profile, forming an $\ExB$ staircase\cite{kosuga2014,r4,r5,r6}. These shear layers act as permeable transport barriers. Their radial organization can modify the relation between heat flux and temperature gradient\cite{r5,r6,peeters2016}.

The fast-ion-regulated enhancement (FIRE) mode observed in KSTAR links fast-ion-induced stabilization of core turbulence to improved confinement, stably maintaining an ion temperature of 10 keV, which is favorable for fusion\cite{r7,r8}. Experimental analyses have associated ion internal-transport-barrier formation in FIRE mode with a transport bifurcation\cite{r8}, suggesting access to a reduced-transport regime in the ion channel. Local nonlinear gyrokinetic simulations based on FIRE-mode conditions identified dilution effects, including a reduced thermal-main-ion density fraction and a modified main-ion density gradient, as the main mechanism of turbulence suppression\cite{r9}. Zonal-flow amplitudes and shearing rates also increased with fast ions, although this increase was not identified as the principal cause of the transport reduction under those conditions\cite{r9}. Reduced theories predict that fast ions can facilitate zonal-flow generation\cite{r10}, modify drift-wave--zonal-flow self-regulation\cite{r11}, and reduce the critical shear for staircase stability\cite{r12}. However, the effects of fast ions on the global radial organization of zonal flows and their associated shear layers, together with the accompanying ion heat-flux response during profile relaxation under FIRE-mode conditions, remain less well characterized.

Especially when zonal flows and temperature profiles develop radial structure spontaneously, a local flux--gradient relation may not fully characterize the transport. Dif-Pradalier et al.\cite{r5} showed that substantially different heat fluxes can occur at similar ambient temperature gradients when the degree of staircase organization differs. This non-local heat transport behavior suggests that local simulation alone may be insufficient when discussing enhanced confinement in FIRE mode. 

In this study, we perform nonlinear global gyrokinetic simulations of FIRE-mode-like plasmas using a high-fidelity code GENE\cite{r19,r20}, to investigate the radial organization of zonal flows and shear layers associated with fast ions. For the reference initial main-ion temperature profile, we compare the spatiotemporal evolution of zonal-flow velocity and shearing rate and the time-averaged radial profiles of the main-ion temperature gradient and heat flux between simulations with and without fast ions. The response to the initial main-ion temperature gradient is then compared between simulations with and without fast ions through a five-point gradient scan using a signed heat-flux integral evaluated over a common radial and temporal interval.

The remainder of this paper is organized as follows. Section 2 describes the physical model and initial conditions (Section 2.1), the numerical setup and normalization (Section 2.2), and the flow and heat-transport diagnostics (Section 2.3). Section 3 presents the results, focusing on zonal flows, temperature-gradient corrugations, and radial heat-flux profiles (Section 3.1), followed by the dependence of the heat-flux response on the initial main-ion temperature gradient (Section 3.2). Section 3.3 discusses the physical interpretation of these results and the scope of the present simulations. Finally, Section 4 summarizes the main conclusions.

\section{Physical model and numerical methods}\label{sec2}

\subsection{Physical model and initial conditions}

We use the radially global version of the Gyrokinetic Electromagnetic Numerical Experiment (GENE) code\cite{r19,r20}, an open-source framework for simulating plasma microturbulence. It resolves fluctuations on gyroradius scales and calculates the resulting turbulent transport coefficients in magnetized fusion plasmas\cite{geneweb}. Previous GENE studies have examined electromagnetic stabilization of tokamak microturbulence at high $\beta$\cite{citrin2015} and self-organized criticality in ion-temperature-gradient (ITG) mode-driven turbulence\cite{mavridis2014}. Global GENE simulations have also been used to investigate fast-ion-induced core transport barriers\cite{r14}. In this study, we exploit this capability of GENE to capture high-$\beta$ effects and the global behavior of ITG turbulence in the presence of fast ions to investigate nonlocal turbulence and heat transport in FIRE-mode plasmas, including phenomena such as $\ExB$ staircase formation.

\begin{figure}[!t]
\centering
\includegraphics[width=\textwidth]{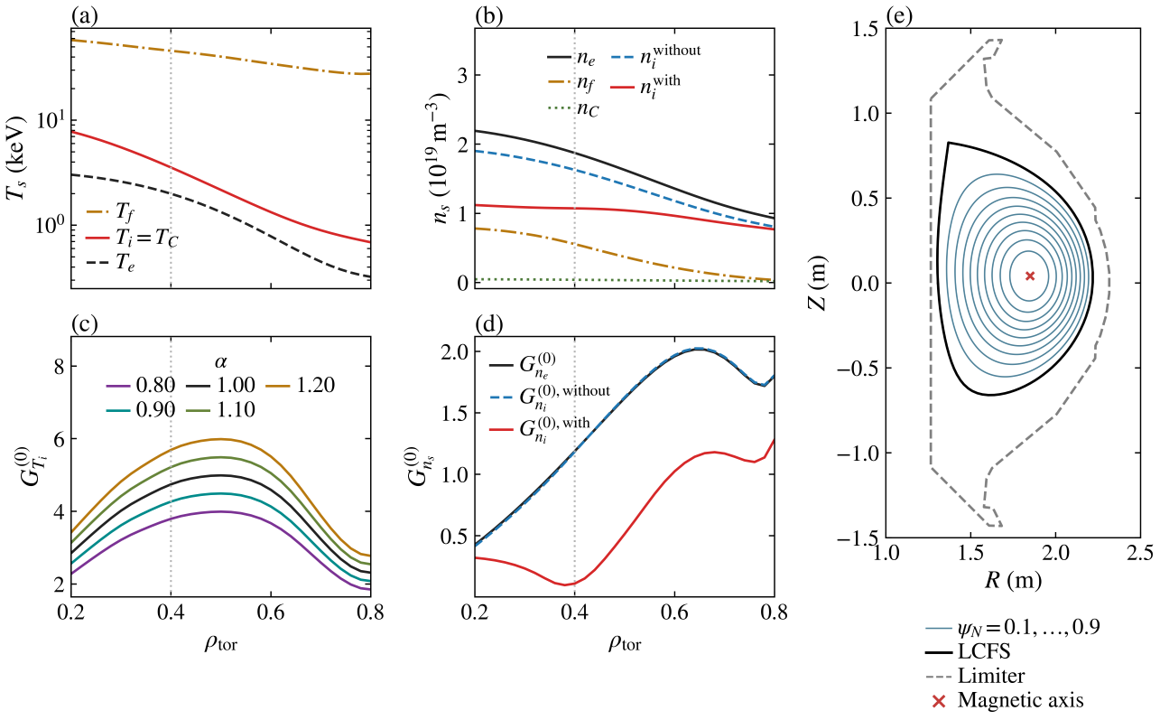}
\caption{Initial profiles and magnetic equilibrium for KSTAR discharge \#22663. The profiles in (a)--(d) correspond to $t=5.350\,\mathrm{s}$. (a) Species temperatures at the reference gradient ($\alpha=1$); (b) densities with and without fast ions; (c) initial main-ion logarithmic temperature gradients for the five scale factors; (d) electron and main-ion logarithmic density gradients; and (e) poloidal cross-section of the EFIT equilibrium used in the simulations, showing normalized poloidal-flux contours, the last closed flux surface (LCFS), the limiter and the magnetic axis. The temperature axis in (a) is logarithmic, and the carbon and main-ion temperatures are equal. In (b,d), the red solid and blue dashed curves show main-ion profiles with and without fast ions, respectively. The initial temperature gradients in (c) are identical for the two fast-ion conditions. Vertical dotted lines in (a)--(d) mark $\rho_a=0.4$.}
\label{fig:profiles}
\end{figure}

FIRE mode is characterized by the formation of a core ion transport barrier associated with a high fast-ion fraction, which can be sustained stably without external control and maintain ion temperatures $>10$ keV for tens of seconds \cite{r7,r8}. In this study, we used KSTAR FIRE-mode discharge \#22663 at $t=5.35\,\mathrm{s}$ for the initial density and temperature profiles. The same magnetic equilibrium is used in all simulations. Here, the electron temperature and density are measured using Thomson scattering \cite{Lee_FED_2023_kstar_thomson}, while the ion temperature is measured using charge-exchange spectroscopy \cite{Lee_FED_2026_kstar_ces}. The fast ion profiles are obtained using NUBEAM \cite{Pankin_CPC_2004_nubeam}, and the magnetic equilibrium is reconstructed using EFIT \cite{Lau_NF_1985_efit}. In the simulation using GENE, electrons, thermal deuterium main ions, and fully ionized carbon impurities are treated kinetically. For consideration of fast ions, high-energy deuterium ions are included as an additional kinetic species. The background distribution of each species is approximated as Maxwellian. The electromagnetic model includes the perturbed electrostatic potential $\varphi_1$ and parallel vector potential $\delta A_\parallel$, while neglecting parallel magnetic-field fluctuations $\delta B_\parallel$ and collisional effects. Externally imposed equilibrium $\ExB$ shear and parallel flow shear are excluded to focus on the self-consistent evolution of turbulence and zonal flows driven by fast ions. No profile-maintaining heat or particle sources are applied, so the simulations describe relaxation from the prescribed initial state. The radial coordinate $\rhot$ denotes the normalized toroidal-flux radius. Figure~\ref{fig:profiles}(a,b) shows the initial temperature and density profiles.

We examine how main-ion heat transport varies with the initial temperature gradient and how fast ions affect this response. Five initial main-ion temperature gradients are considered with and without fast ions, giving ten simulations in total. Starting from a reference profile with scale factor $\alpha=1$ ($T_{i,1}(\rhot)$), we use $\alpha=0.80$, $0.90$, $1.00$, $1.10$ and $1.20$. The temperature at the reference radius $\rho_a=0.4$ is held fixed by applying
\begin{equation}
T_{i,\alpha}(\rhot)=T_{i,1}(\rho_a)
\left[
\frac{T_{i,1}(\rhot)}
{T_{i,1}(\rho_a)}
\right]^\alpha .
\label{eq:scan}
\end{equation}
The carbon temperature profile is set equal to the main-ion temperature profile. The logarithmic gradients of the main-ion temperature and species-$s$ density are defined as
$\GT\equiv-\partial\ln T_i/\partial\rhot$ and
$G_{n_s}\equiv-\partial\ln n_s/\partial\rhot$, respectively.
A superscript $(0)$ denotes the initial value. Equation~\eqref{eq:scan} scales $\GT^{(0)}$ by $\alpha$ without changing the temperature at $\rho_a$ [Figure~\ref{fig:profiles}(c)].

For the comparison with and without fast ions, the electron and carbon density profiles are kept fixed. Quasineutrality is maintained by setting the main-ion densities to
\begin{equation}
n_i^{\mathrm{with}}=n_e-n_f-6n_C,
\qquad
n_i^{\mathrm{without}}=n_e-6n_C,
\label{eq:dilution}
\end{equation}
where $n_e$, $n_f$ and $n_C$ denote electron, fast-ion and carbon densities. The superscripts distinguish simulations with and without fast ions. The two conditions consequently differ in both the main-ion density fraction and its gradient [Figure~\ref{fig:profiles}(b,d)].

\subsection{Numerical setup}

The magnetic geometry is constructed from an EFIT gEQDSK equilibrium using the TRACER-EFIT interface \cite{Lau_NF_1985_efit}. The poloidal cross-section of this equilibrium is shown in Figure 1(e). The radial computational domain is $0.1\leq\rhot\leq0.9$, with Dirichlet conditions at both boundaries. A buffer region spanning $10\%$ of the total radial domain width is used at each boundary. Krook damping is applied in these buffers with a normalized boundary coefficient of $2.0$ and radial exponent of $4$.

The phase space is discretized in the radial direction, binormal Fourier modes, the field-aligned direction, parallel velocity and magnetic moment\cite{r15,r16}. The corresponding resolution is \((N_x,N_{k_y},N_z,N_{v_\parallel},N_\mu)=(256,32,48,32,16)\). All simulations use the same numerical settings and reference quantities. Time, velocity and length are normalized using $\tN=t c_{\mathrm{ref}}/L_{\mathrm{ref}}$, $c_{\mathrm{ref}}=\sqrt{T_{\mathrm{ref}}/m_{\mathrm{ref}}}$ and
$\rho_{\mathrm{ref}}=c_{\mathrm{ref}}/\Omega_{\mathrm{ref}}$,
where $L_{\mathrm{ref}}$, $T_{\mathrm{ref}}$, $m_{\mathrm{ref}}$ and $B_{\mathrm{ref}}$ are the reference length, temperature, mass and magnetic field, and $\Omega_{\mathrm{ref}}$ is the reference cyclotron angular frequency. The minimum nonzero normalized binormal wavenumber is $k_{y,\min}\rho_{\mathrm{ref}}=0.043154129$. The velocity-space domain is $-3.7\leq v_\parallel/v_{Ts,0}\leq3.7$ and $0\leq\mu B_{\mathrm{ref}}/T_{s,0}\leq13$, where $T_{s,0}$ is the initial temperature of species $s$ at $\rhot=0.5$ and $v_{Ts,0}=\sqrt{2T_{s,0}/m_s}$.

Fourth-order hyperdiffusion is applied in the radial and field-aligned directions to suppress grid-scale numerical noise. The GENE input parameters are \texttt{hyp\_x = 0.1} and \texttt{hyp\_z = -0.5}. The negative value of \texttt{hyp\_z} selects the grid-spacing-dependent prescription for parallel numerical dissipation\cite{r16}. The parameter files used for the simulations are available in the accompanying
repository\cite{r24}.

\subsection{Flow and heat-transport diagnostics}

The normalized zonal $\ExB$ velocity and $\ExB$ shearing rate are defined by
\begin{equation}
\VZ=
\frac{1}{C_{xy}}
\frac{\partial\pZ}{\partial\xi},
\label{eq:velocity}
\end{equation}
and
\begin{equation}
\wZ=
\frac{1}{C_{xy}}
\frac{\partial^2\pZ}{\partial\xi^2}.
\label{eq:shear}
\end{equation}
Here, $\pZ$ is the flux-surface average of the $k_y=0$ component of the normalized perturbed potential, $\widehat{\varphi}=e\varphi_1L_{\mathrm{ref}}/(T_{\mathrm{ref}}\rho_{\mathrm{ref}})$.
The differentiation coordinate is $\xi=\rhot/\rho_*$, where
$\rho_*=\rho_{\mathrm{ref}}/a$ and $a$ is the minor radius.
The dimensionless geometric coefficient $C_{xy}$ is determined by the magnetic geometry.
The velocity and shearing rate are normalized as
$\VZ=V_{E,y}^{\mathrm{ZF}}/(\varepsilon_{\mathrm{ref}}c_{\mathrm{ref}})$ and
$\wZ=\omega_{E\times B}/(c_{\mathrm{ref}}/L_{\mathrm{ref}})$,
respectively, with
$\varepsilon_{\mathrm{ref}}=\rho_{\mathrm{ref}}/L_{\mathrm{ref}}$.

The main-ion heat-flux diagnostic $Q_i$ is the normalized turbulent radial heat flux, including electrostatic and electromagnetic contributions. The separately evaluated neoclassical contribution is not included. The gyro-Bohm normalization is \(Q_{\mathrm{gB}}=n_{\mathrm{ref}}T_{\mathrm{ref}}c_{\mathrm{ref}}
(\rho_{\mathrm{ref}}/L_{\mathrm{ref}})^2\), where $n_{\mathrm{ref}}$ is the reference density and the other reference quantities are defined above. The same normalization is used in all simulations. Positive and negative fluxes denote outward and inward transport, respectively.

Radial profiles of the temperature gradient, flow and heat flux are time-averaged over a sufficiently saturated time region,
$15\leq\tN\leq24$, where $\langle\cdot\rangle_t$ denotes the time average over this interval. The response to the initial temperature gradient is compared using the signed integral
\begin{equation}
\Ii=
\int_{15}^{24}
\int_{0.32}^{0.60}
Q_i(\rhot,\tN)
\,\mathrm{d}\rhot
\,\mathrm{d}\tN .
\label{eq:integral}
\end{equation}
Here, the $0.32\leq\rhot\leq0.60$ region was selected as the core region where the ion temperature gradient becomes steep in the experiment, as shown in Figure~\ref{fig:profiles}. Oppositely directed contributions cancel in this signed integral. In this study, the quantity $\Ii$ is used to compare the finite-time transport response.

\section{Results and discussion}

\subsection{Zonal flows and temperature-gradient corrugations}

\begin{figure}[!t]
\centering
\includegraphics[width=15cm]{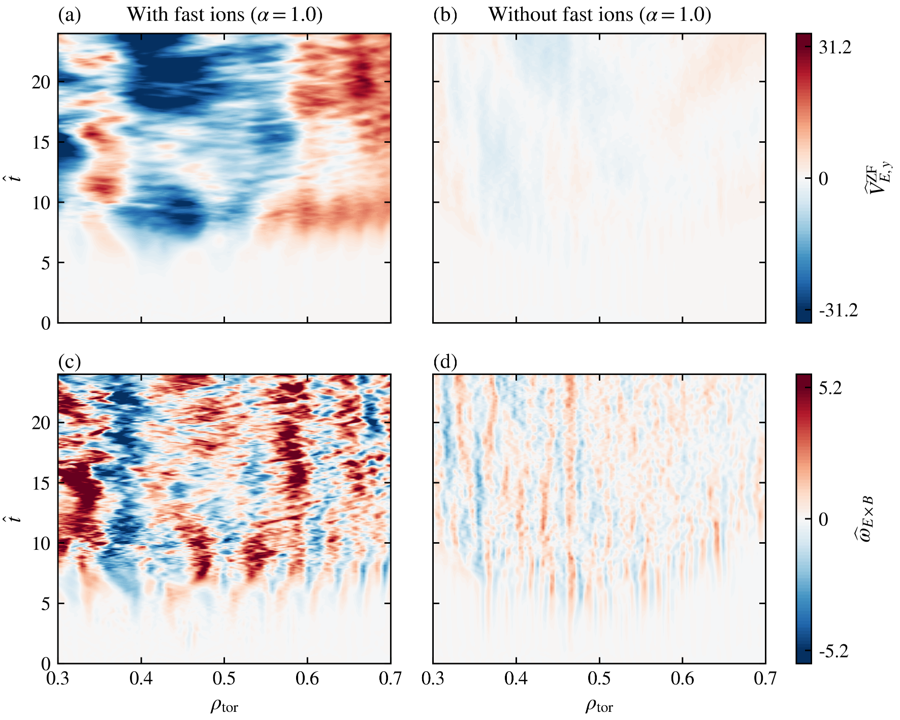}
\caption{Normalized zonal $\ExB$ velocity $\VZ$ (a,b) and $\ExB$ shearing rate $\wZ$ (c,d) at the reference gradient ($\alpha=1$). Left: with fast ions; right: without fast ions. Color limits are set by the 97.5th percentile of the combined absolute values from the two simulations.}
\label{fig:flow}
\end{figure}

Previous local gyrokinetic simulations of KSTAR FIRE mode identified main-ion density dilution induced by fast ions as the primary mechanism of turbulence suppression\cite{r9}. Under ASDEX Upgrade conditions, gyrokinetic simulations have also reported core transport barriers associated with wave--particle resonant interactions between fast ions and ion-scale turbulence\cite{r14}, with the resonant effect depending on the fast-ion temperature and density profiles. Here, we focus on how these direct and indirect fast-ion effects on drift waves and zonal flows manifest at the mesoscale. Figure~\ref{fig:flow} compares the spatiotemporal distributions of zonal-flow velocity and shearing rate at the reference gradient, $\alpha=1$, obtained from the global gyrokinetic simulations described in Section \ref{sec2}.

With fast ions [Figures~\ref{fig:flow}(a, c)], oppositely directed flows form at different radii, and large-amplitude flows and localized shear layers persist into the later part of the analysis interval ($\tN\gtrsim10$). Flows and shear also fluctuate without fast ions, but with smaller amplitudes, as shown in Figures~\ref{fig:flow}(b, d). Within the displayed radial and temporal interval, the maximum absolute values of $\VZ$ are $48.2$ with fast ions and $6.45$ without fast ions. The corresponding maximum absolute values of $\wZ$ are $15.0$ and $3.13$. In the fast-ion case at the reference gradient, elongated, brush-like fine structures are visible between the pronounced zonal jets. Their appearance is suggestive of avalanche-like radial propagation \cite{r5}.

\begin{figure}[!t]
\centering
\includegraphics[width=15cm]{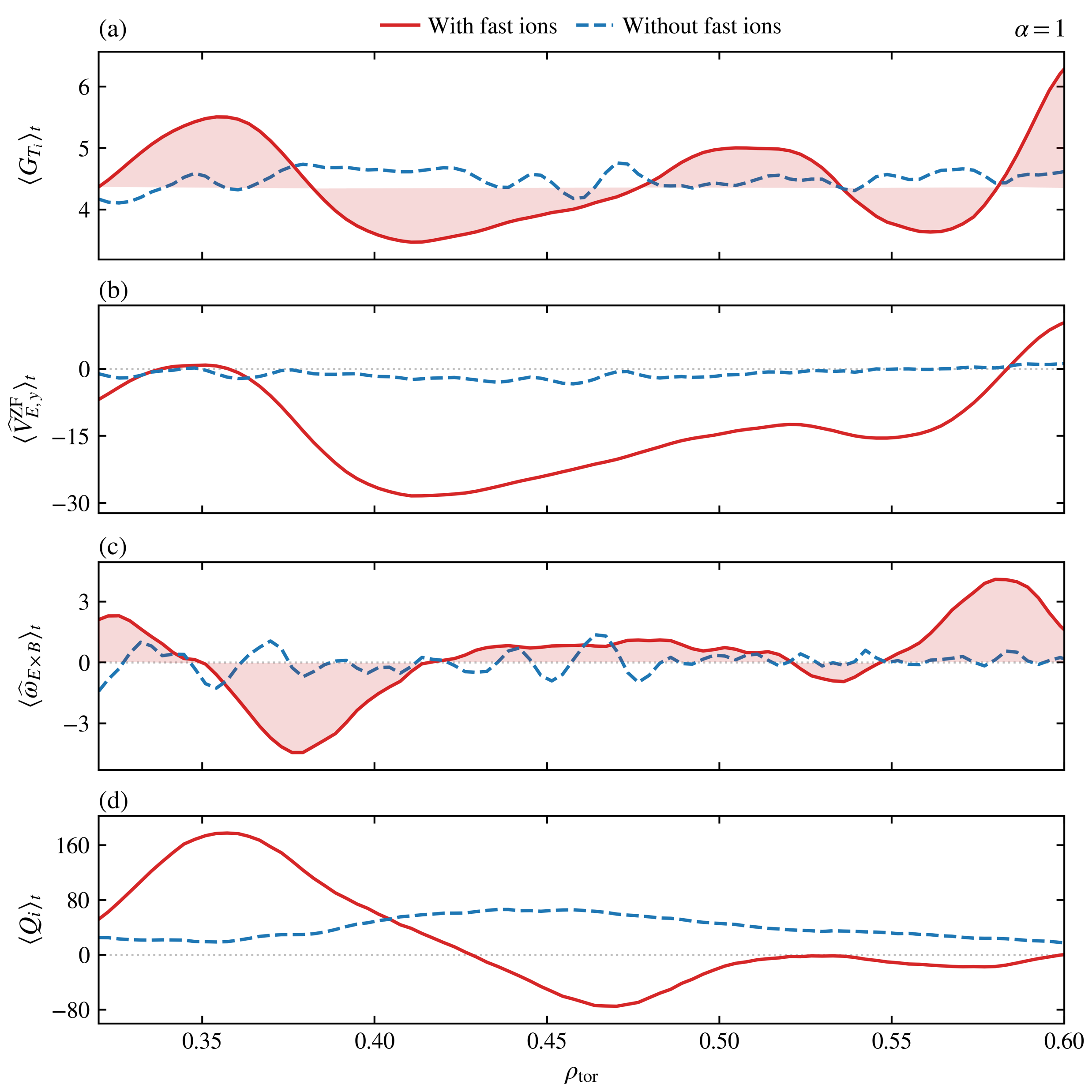}
\caption{Time-averaged radial profiles at the reference gradient ($\alpha=1$): (a) main-ion logarithmic temperature gradient $\avg{\GT}$; (b) normalized zonal $\ExB$ velocity $\avg{\VZ}$; (c) normalized $\ExB$ shearing rate $\avg{\wZ}$; and (d) main-ion heat flux $\avg{Q_i}$. The averaging interval is $15\leq\tN\leq24$ and the radial interval is $0.32\leq\rhot\leq0.60$. Red solid and blue dashed curves denote simulations with and without fast ions, respectively. Horizontal dotted lines in (b--d) indicate zero.}
\label{fig:mean}
\end{figure}

Figure~\ref{fig:mean}(a--c) compares the temperature gradient, zonal-flow velocity and shearing rate averaged over $15\leq\tN\leq24$. With fast ions, regions of high and low main-ion temperature gradient alternate radially, accompanied by strong radial variations in the zonal-flow velocity and shearing rate. Without fast ions, the gradient corrugations and flow amplitudes are weaker. The coexistence of persistent shear layers and temperature-gradient corrugations is qualitatively consistent with the spatial structure of $\ExB$ staircases\cite{r4,r5}.

The radial heat-flux profiles differ in both magnitude and spatial structure [Figure~\ref{fig:mean}(d)]. With fast ions, the time-averaged normalized heat flux reaches an outward maximum of approximately 180 near $\rhot\simeq0.35$, changes sign near $\rhot\simeq0.43$, and reaches a minimum of approximately $-75$ near $\rhot\simeq0.47$. Without fast ions, the flux remains outward throughout the radial interval examined, with a maximum of approximately 65 near $\rhot\simeq0.45$. The outward peak with fast ions is thus approximately 2.8 times that without fast ions and is located farther inward. This behavior is analogous to the overshoot of limit-cycle oscillations beyond the static equilibrium level in a conventional predator–prey system \cite{Huang_NF_2026_predator_prey}, suggesting that the predator–prey interaction between zonal flows and turbulence may manifest not only on temporal scales but also through spatial organization.

In the simulation with fast ions, inward flux occurs even where the mean logarithmic temperature gradient is positive. The larger outward peak and the coexistence of outward and inward fluxes indicate that fast ions modify the radial distribution of heat transport rather than uniformly reducing the flux across the examined interval. The inward heat-flux component may be relevant to ion confinement and transport-barrier formation through radial redistribution of ion heat. Its contribution to the plasma energy balance requires further investigation.

\begin{figure}[!t]
\centering
\includegraphics[width=15cm]{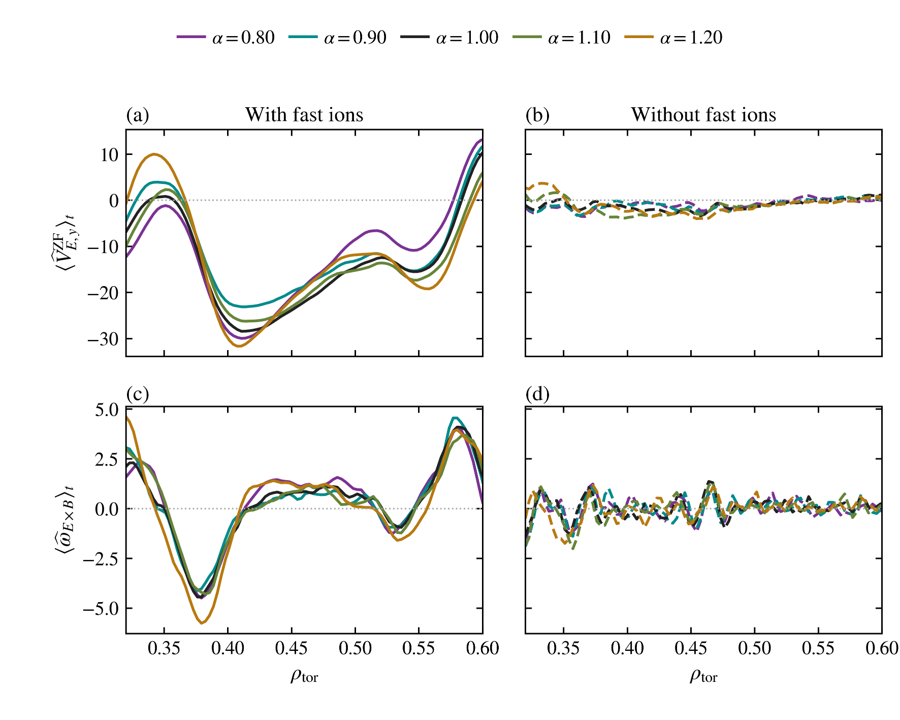}
\caption{Time-averaged normalized zonal $\ExB$ velocity $\avg{\VZ}$ (a,b) and shearing rate $\avg{\wZ}$ (c,d) for the five initial main-ion temperature gradients. Left: with fast ions; right: without fast ions. Colors denote $\alpha$. The averaging interval is $15\leq\tN\leq24$ and the displayed radial interval is $0.32\leq\rhot\leq0.60$. Horizontal dotted lines indicate zero.}
\label{fig:zonal_scan}
\end{figure}

Persistent shear layers are also observed with fast ions at $\alpha=0.80$, $0.90$, $1.10$ and $1.20$ (supplementary Figures S1--S4). The zonal-flow velocity and shearing rate averaged over $15\leq\tN\leq24$ have larger overall amplitudes with fast ions across all five initial gradients [Figure~\ref{fig:zonal_scan}]. In the fast-ion cases, a negative shear trough near $\rhot\simeq0.38$ and a positive peak near $\rhot\simeq0.58$ are common across the scan, although their amplitudes and detailed radial shapes vary with $\alpha$.

Across the five fast-ion cases, the minimum time-averaged normalized zonal-flow velocity ranges approximately from $-32$ to $-23$, whereas the corresponding profiles without fast ions remain much closer to zero. The negative shear extrema range approximately from $-6$ to $-4$, while the outer positive extrema range approximately from $3.5$ to $4.5$. The dependence of the shear amplitude on the initial temperature gradient varies with radius: the deepest negative trough occurs at $\alpha=1.20$, whereas the largest outer positive peak occurs at $\alpha=0.90$. Thus, increasing the initial temperature gradient does not produce a uniform amplification of the radial shear profile.

Here, the differences in zonal-flow velocity and shearing rate across the initial temperature-gradient cases shown in Figure~\ref{fig:zonal_scan} may appear relatively small by visual inspection. However, because these quantities exhibit spatially corrugated structures with alternating positive and negative regions, even modest local differences can lead to significant changes in their spatial integrals, as will be discussed in detail in Section \ref{sec3.2}.

\FloatBarrier

\subsection{Heat-flux response to the initial temperature gradient}\label{sec3.2}

\begin{figure}[!t]
\centering
\includegraphics[width=\textwidth]{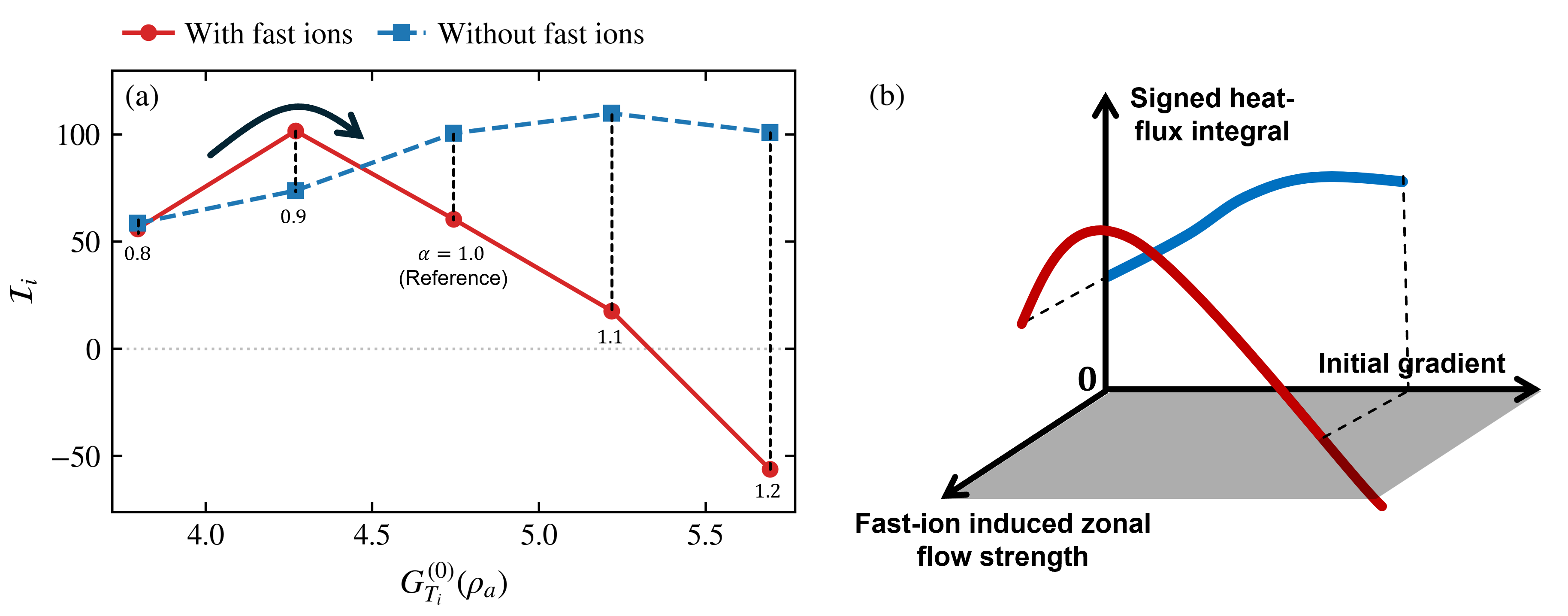}
\caption{(a) Signed radial and temporal heat-flux integral $\Ii$ versus the initial main-ion logarithmic temperature gradient $\GT^{(0)}(\rho_a)$, evaluated at $\rho_a=0.4$. From left to right, the points on each curve correspond to $\alpha=0.80$, $0.90$, $1.00$, $1.10$ and $1.20$. Red circles with a solid line and blue squares with a dashed line denote simulations with and without fast ions, respectively. Integration is performed over $0.32\leq\rhot\leq0.60$ and $15\leq\tN\leq24$, retaining the sign of the heat flux [equation~\eqref{eq:integral}]. Lines connect the simulated cases. (b) Schematic representation of the different signed heat-flux responses to the initial main-ion temperature gradient for the two fast-ion conditions shown in (a). The curves illustrate the observed trends and do not represent additional simulations.}
\label{fig:scan}
\end{figure}

Although the simulations showed that heat transport varies with radial position, experimentally observed mesoscale behaviors such as transport barriers are determined by the combined dynamics over a finite radial region rather than at a single location. Therefore, in this section, we further analyze the radial profiles shown in Figures 3 and 4 using spatially integrated quantities, as defined in Equation (\ref{eq:integral}). Figure~\ref{fig:scan} shows the integrated heat flux $\Ii$ with respect to the initial temperature gradient.

In the absence of fast ions, the heat flux mostly increases with the ion-temperature gradient, consistent with the conventional diffusive intuition based on Fick's law. Under external heating, this behavior tends to limit further growth of the temperature gradient because the increased outward heat transport balances the applied heating. In the presence of fast ions, however, an interesting behavior emerges. The heat flux initially increases with the temperature gradient, consistent with the conventional diffusive trend, but beyond $\alpha = 0.9$, it instead decreases as the temperature gradient is further increased. Such a relation resembles the transition-like curves often observed in association with core transport barrier formation \cite{Diamond_PRL_1997}, as illustrated in Figure~\ref{fig:scan}(b). Notably, the FIRE-mode discharge $\alpha = 1.0$, a bifurcated regime in KSTAR, lies within this negative-correlation region, which may be relevant to the observed confinement improvement. In particular, we note that this confinement improvement arises from the integrated effects of the nonlocal behaviors discussed in Section 3.1. Previous studies \cite{r7,r8} have already attributed the confinement improvement in KSTAR FIRE mode primarily to local effects, including main-ion density dilution and local fast-ion stabilization. However, these mechanisms mainly describe a continuous improvement in performance from conventional L-mode as fast-ion effects become stronger, and do not fully explain the clearly bifurcated regime observed experimentally. In contrast, the turnover of the heat-flux curve in Figure~\ref{fig:scan} may provide a clue to the origin of this bifurcation.

Dif-Pradalier et al.\cite{r5} discussed how zonal-flow strength can act as an additional control parameter in the flux--gradient relation, modifying its nonlinear response or potentially inducing transport bifurcations. In the present simulations, the fast-ion cases exhibit stronger time-averaged zonal flows and shearing rates across the initial-gradient scan [Figure~\ref{fig:zonal_scan}], together with a different dependence of the signed heat-flux integral on the initial gradient [Figure~\ref{fig:scan}(a)]. This difference is schematically illustrated in Figure~\ref{fig:scan}(b). Together with the inward heat flux observed with $\avg{\GT}>0$ [Figure~\ref{fig:mean}(a,d)], these findings suggest that the local temperature gradient alone may not fully characterize ion heat transport in the fast-ion cases.

In the fast-ion case at $\alpha=1.20$, $\Ii<0$, indicating that the inward contribution exceeds the outward contribution over the selected radial and temporal interval. Because $\Ii$ is evaluated over a finite interval during profile relaxation, this negative value does not represent steady-state inward transport or the total energy loss of the plasma. Nonlocal transport associated with the global flow organization is one possible mechanism contributing to the inward heat-flux component. At the experimentally based reference profile, this inward contribution partly offsets the outward contribution in the signed integral and may help account for its smaller value with fast ions. The underlying mechanism and its possible relevance to improved ion confinement in FIRE mode require further investigation.


\subsection{Physical interpretations}


Nonlinear global gyrokinetic simulations inherently include the combined effects of multiple mechanisms and chaotic dynamics, which limits a clear identification of why fast ions enhance zonal flows and shearing rates and, ultimately, promote staircase-like structures. Fortunately, existing theory-based reduced models provide qualitative dependencies that clarify the role of fast ions in several specific mechanisms. In this subsection, we evaluate these dependencies using parameter values representative of FIRE-mode conditions and relate the predictions of the reduced models to the results of our high-fidelity simulations.

At $\rho_{\mathrm{tor}}=0.4$ in the experimentally based reference initial profile, $T_f/T_e\simeq22.94$, the fast-ion charge-density fraction is $f_h=Z_fn_{f0}/n_{e0}\simeq0.30$ with $Z_f=1$, and the density-gradient scale-length ratio is $\eta_n=L_{ne}/L_{ni}\simeq0.090$, where $L_{ns}=n_{s0}/|\nabla n_{s0}|$. We use these parameters to compare the observed zonal-flow development and shear-layer persistence with reduced-model predictions.

Reduced models predict that fast-ion dilution weakens drift-wave dispersion and lowers the amplitude threshold for zonal-flow excitation \cite{r10,r11}. The coupled drift-wave--zonal-flow model also predicts reduced saturated turbulence and transport in the weak-turbulence regime \cite{r11}. In the non-resonant, long-wavelength treatment, the modulational growth rate satisfies \cite{r12}
\[
\Gamma^2=\gamma_{\mathrm{mod}}^2-\Delta_{\mathrm{mm}}^2,
\]
where $\gamma_{\mathrm{mod}}$ represents the zonal-flow drive and $\Delta_{\mathrm{mm}}$ represents frequency detuning. At fixed drift-wave intensity and spectral shape,
\[
\gamma_{\mathrm{mod}}^2=(1-f_h)\gamma_{\mathrm{mod}(0)}^2,
\qquad
\Delta_{\mathrm{mm}}^2\simeq
(1-f_h)^4\eta_n^2\Delta_{\mathrm{mm}(0)}^2.
\]
Here, the subscript $(0)$ denotes the idealized fast-ion-free model with $\eta_n=1$, distinct from the superscript $(0)$ used for the initial simulation profiles. With the parameters above, the squared drive is reduced by a factor of $0.70$, whereas the squared detuning is reduced by approximately $1.9\times10^{-3}$. This stronger reduction of detuning lowers the model's excitation threshold, offering a possible interpretation of the pronounced zonal flows in our simulations.

Wave trapping in zonal-flow shear layers has been proposed as a mechanism contributing to staircase formation \cite{r23}. For the initially saturated drift-wave--zonal-flow system in the moderately narrowband regime, Choi et al. derive a critical shearing rate for intrinsic staircase stability \cite{r12}. Applying the saturation condition $\gamma_{\mathrm{mod}}^2=\Delta_{\mathrm{mm}}^2$ and assuming an unchanged shape of the wave-population distribution gives
\[
\frac{\omega_{E\times B}^{\mathrm{crit}}}
{\omega_{E\times B(0)}^{\mathrm{crit}}}
=(1-f_h)^8\eta_n^4
\simeq3.8\times10^{-6}.
\]
This formal reduced-model estimate indicates a lower shear threshold against intrinsic structural collapse and is qualitatively consistent with the persistent shear layers observed in our global gyrokinetic simulation with fast ions. Here, given the idealized assumptions of the reduced model, this estimate should be interpreted qualitatively rather than as a quantitative prediction for the FIRE-mode plasma. 



\section{Conclusions}

Nonlinear and global gyrokinetic simulations based on KSTAR FIRE-mode conditions were used to examine fast-ion effects on zonal flows and ion heat transport during profile relaxation. At the ion-temperature-gradient condition observed in the experiment, fast ions develop larger zonal-flow amplitudes and shearing rates than the simulation without fast ions. Persistent shear layers coexist with pronounced main-ion temperature-gradient corrugations, qualitatively consistent with the spatial structure of $E\times B$ staircases. Larger zonal-flow and shearing-rate amplitudes with persistent shear layers are also observed at the other four temperature gradients examined. When fast ions are absent, the average main-ion heat flux remains outward throughout the radial interval, whereas inward and outward fluxes coexist at different radii when fast ions are present. 

The coexistence and radial interchange of inward and outward heat fluxes cause the transport response to deviate from the conventional Fick's law trend, in which the heat flux is expected to increase with the temperature gradient. In numerical experiments with varied initial temperature gradients, we find that the main-ion heat flux can instead decrease as the temperature gradient increases in the presence of fast ions. This produces a non-monotonic, transition-like heat-flux response resembling the flux--gradient behavior associated with core transport-barrier formation. Such behavior may contribute to the experimental observation that FIRE mode in KSTAR represents a bifurcated regime rather than a continuous performance improvement from conventional L-mode. The finding from global gyrokinetic simulations that fast ions can promote nonlocal $E\times B$ staircase formation, thereby driving the heat-flux response away from the Fick’s law trend and potentially enabling a transport bifurcation, provides a distinctive global feature that is inaccessible to conventional local flux-tube simulations.

These results suggest that fast ions may modify the conventional Fick’s-law-based paradigm of mesoscale heat transport. This possibility is particularly relevant to reactor-grade plasmas such as ITER, where fusion-born fast alpha particles are abundant and their influence on fuel-ion confinement can directly affect fusion power and gain. Global gyrokinetic simulations of fast-ion effects under such reactor-relevant conditions will therefore be an important subject for future study.



\section*{Acknowledgements}

This research was supported by the Chung-Ang University Graduate Research Scholarship in 2024. This work was also supported by a National Research Foundation of Korea (NRF) grant funded by the Korea government (MSIT) (Grant No. RS-2024-00346024). Computing resources were provided on the KFE computer, KAIROS, funded by the Ministry of Science and ICT of the Republic of Korea (No. KFE-EN2641-12). The authors used ChatGPT (OpenAI) to assist with literature searches and explanations, discussion of simulation settings and physical interpretations, development and revision of data-analysis and plotting scripts, and drafting, translation and editing of the manuscript. The authors take responsibility for the scientific content and the final manuscript.

\end{document}